\documentclass[final,5p,times,twocolumn]{elsarticle}

\usepackage{graphicx}

\usepackage{amssymb}

\begin{document}

\begin{frontmatter}



\title{Properties of resonant states in $^{18}$Ne relevant to key $^{14}$O($\alpha$,$p$)$^{17}$F breakout reaction in type I x-ray bursts}


\author[1,2]{J.~Hu}
\author[1]{J.J.~He\corref{cor1}}
\ead{jianjunhe@impcas.ac.cn}
\author[3,4]{A.~Parikh\corref{cor1}}
\ead{anuj.r.parikh@upc.edu}
\cortext[cor1]{Corresponding authors.}
\author[1,5]{S.W.~Xu}
\author[2]{H.~Yamaguchi}
\author[2]{D.~Kahl}
\author[1]{P.~Ma}
\author[6]{J.~Su}
\author[7]{H.W.~Wang}
\author[2]{T.~Nakao}
\author[8]{Y.~Wakabayashi}
\author[9]{T.~Teranishi}
\author[10]{K.I.~Hahn}
\author[11]{J.Y.~Moon}
\author[11]{H.S.~Jung}
\author[12]{T.~Hashimoto}
\author[13]{A.A.~Chen}
\author[13]{D.~Irvine}
\author[11]{C.S.~Lee}
\author[1,8]{S.~Kubono}

\address[1]{Key Laboratory of High Precision Nuclear Spectroscopy and Center for Nuclear Matter Science, Institute of Modern Physics, Chinese Academy of Sciences, Lanzhou 730000, China}%
\address[2]{Center for Nuclear Study (CNS), the University of Tokyo, Wako Branch at RIKEN, 2-1 Hirosawa, Wako, Saitama 351-0198, Japan}
\address[3]{Departament de F\'{\i}sica i Enginyeria Nuclear, EUETIB, Universitat Polit\`{e}cnica de Catalunya, Barcelona E-08036, Spain}
\address[4]{Institut d'Estudis Espacials de Catalunya, Barcelona E-08034, Spain}
\address[5]{University of Chinese Academy of Sciences, Beijing 100049, China}
\address[6]{China Institute of Atomic Energy (CIAE), P.O. Box 275(46), Beijing 102413, China}
\address[7]{Shanghai Institute of Applied Physics (SINAP), Chinese Academy of Sciences (CAS), Shanghai 201800, China}
\address[8]{RIKEN Nishina Center, 2-1 Hirosawa, Wako, Saitama 351-0198, Japan}
\address[9]{Department of Physics, Kyushu University, 6-10-1 Hakozaki, Fukuoka 812-8581, Japan}
\address[10]{Department of Science Education, Ewha Womans University, Seoul 120-750, Republic of Korea}
\address[11]{Department of Physics, Chung-Ang University, Seoul 156-756, Republic of Korea}
\address[12]{Research Center for Nuclear Physics (RCNP), Osaka University, 10-1 Mihogaoka, Ibaraki, Osaka, 567-0047, Japan}
\address[13]{Department of Physics \& Astronomy, McMaster University, Hamilton, Ontario L8S 4M1, Canada}

\begin{abstract}
The $^{14}$O($\alpha$,$p$)$^{17}$F reaction is one of the key reactions involved in the breakout from the hot-CNO cycle to the
rp-process in type I x-ray bursts. The resonant properties in the compound nucleus $^{18}$Ne have been investigated through
resonant elastic scattering of $^{17}$F+$p$. The radioactive $^{17}$F beam was separated by the CNS Radioactive Ion Beam separator (CRIB) and
bombarded a thick H$_2$ gas target at 3.6 MeV/nucleon. The recoiling light particles were measured by using three ${\Delta}$E-E
silicon telescopes at laboratory angles of $\theta$$_{lab}$$\approx$3$^\circ$, 10$^\circ$ and 18$^\circ$, respectively. Five
resonances at $E_{x}$=6.15, 6.28, 6.35, 6.85, and 7.05 MeV were observed in the excitation functions. Based on an $R$-matrix
analysis, $J^{\pi}$=1$^-$ was firmly assigned to the 6.15-MeV state. This state dominates the thermonuclear
$^{14}$O($\alpha$,$p$)$^{17}$F rate below 1 GK. We have also confirmed the existence and spin-parities of three states
between 6.1 and 6.4 MeV. As well, a possible new excited state in $^{18}$Ne was observed at $E_{x}$=6.85$\pm$0.11 MeV and tentatively
assigned as $J$=0. This state could be the analog state of the 6.880 MeV (0$^{-}$) level in the mirror nucleus $^{18}$O, or a bandhead
state (0$^+$) of the six-particle four-hole (6$p$-4$h$) band. A new thermonuclear rate of the $^{14}$O($\alpha$,$p$)$^{17}$F reaction
has been determined, and its astrophysical impact has been examined within the framework of one-zone x-ray burst postprocessing calculations.
\end{abstract}


\end{frontmatter}
Type I x-ray bursts (XRBs) are characterized by sudden dramatic increases in luminosity of roughly 10--100 s in duration, with
peak luminosities of roughly 10$^{38}$ erg/s. The characteristics of XRBs have been surveyed extensively by a number of
space-borne x-ray satellite observatories. More than 90 galactic XRBs have been identified since their initial discovery in 1976.
These recurrent phenomena (on timescales of hours to days) have been the subject of many observational, theoretical and
experimental studies (for reviews see {\it e.g.},~\cite{bib:lew93,bib:str06,bib:par13}). The bursts have been interpreted as
being generated by thermonuclear runaway on the surface of a neutron star that accretes H- and He-rich material from a less
evolved companion star in a close binary system~\cite{bib:woo76,bib:jos77}. The accreted material burns stably through the hot,
$\beta$-limited carbon-nitrogen-oxygen (HCNO)~\cite{bib:wie99} cycles, giving rise to the persistent flux. Once critical
temperatures and densities are achieved, breakout from this region can occur through, {\it e.g.}, $\alpha$-induced reactions on
the nuclei $^{14}$O and $^{15}$O. Through the rapid proton capture process (rp-process)~\cite{bib:wal81,bib:sch98,bib:woo04},
this eventually results in a rapid increase in energy generation (ultimately leading to the XRB) and nucleosynthesis up to
A$\sim$100 mass region~\cite{bib:sch01,bib:elo09}. As one of the trigger reactions, the rate of $^{14}$O($\alpha$,$p$)$^{17}$F
determines, in part, the conditions under which the burst is initiated and thus plays a critical role in understanding burst
conditions~\cite{bib:wie98}.

Contributions from the resonant states dominate the $^{14}$O($\alpha$,$p$)$^{17}$F reaction rate, and therefore the resonant
parameters for the excited states above the $\alpha$ threshold ($Q_\alpha$=5.115 MeV~\cite{bib:wan12}) in the compound nucleus
$^{18}$Ne are required. So far, although our understanding of the reaction rate of $^{14}$O($\alpha$,$p$)$^{17}$F has been greatly
improved via, {\it e.g.}, indirect studies~\cite{bib:hah96,bib:par99,bib:park,bib:gom01,bib:bla03,bib:cern,bib:bar10,bib:he11},
direct study~\cite{bib:not04}, as well as time-reversal studies~\cite{bib:bla01,bib:har99,bib:har02}, most of the required
parameters (such as, $J^{\pi}$ and $\Gamma_\alpha$) have still not been sufficiently well determined over stellar
temperatures achieved in XRBs ($\approx$0.2--2 GK).

In the temperature region below $\sim$1 GK, a state at $E_x$=6.15 MeV (tentatively assigned as 1$^-$, see below) is thought to
dominate the $^{14}$O($\alpha$,$p$)$^{17}$F rate~\cite{bib:hah96}. About twenty-five years ago,
Wiescher {\it et al.}~\cite{bib:wie87} predicted a $J^\pi$=1$^-$ state at $E_x$=6.125 MeV in $^{18}$Ne with a width of
$\Gamma$=$\Gamma_p$=51 keV based on a Thomas-Ehrman shift calculation. Later on, Hahn {\it et al.}~\cite{bib:hah96} observed a
state at $E_x$=6.15$\pm$0.02 MeV through studies of the $^{16}$O($^3$He,$n$)$^{18}$Ne and $^{12}$C($^{12}$C,$^6$He)$^{18}$Ne
reactions. The transferred angular momentum was restricted to be $\ell$$\leq$2 from their measured ($^3$He,$n$) angular
distribution. Based on the Coulomb-shift calculation and prediction of Wiescher {\it et al.}, a $J^\pi$=1$^-$ was tentatively
assigned to this state. G\"{o}mez {\it et al.}~\cite{bib:gom01} studied the resonances in $^{18}$Ne by using the elastic
scattering of $^{17}$F+$p$ and fitted the 6.15-MeV state with 1$^-$ by an $R$-matrix analysis of the excitation function.
However, their 1$^{-}$ assignment was questioned in a later $R$-matrix reanalysis~\cite{bib:arX}. He {\it et al.}~\cite{bib:arX}
thought that this 1$^-$ resonance should behave as a dip-like structure (rather than the peak observed in Ref.~\cite{bib:gom01})
in the excitation function due to the interference. Unfortunately, a recent low-statistics measurement could not resolve this
state~\cite{bib:he11}. Recently, Bardayan {\it et al.}~\cite{bib:bar12} reanalyzed the unpublished elastic-scattering data in
Ref.~\cite{bib:bla03} and also found the expected dip-like structure, however, the statistics were not sufficient to constrain
the parameters of such a resonance. Therefore, three possibilities arise regarding the results presented in Ref.~\cite{bib:gom01}
on the $J^\pi$ of the 6.15 MeV state:
(i) their analysis procedure may be questionable as they needed to reconstruct the excitation functions (above 2.1 MeV)
with some technical treatment since the high-energy protons escaped from two thin Si detectors;
(ii) the peak observed in Ref.~\cite{bib:gom01} may be due to the inelastic scattering contribution~\cite{bib:bar12,bib:gri02},
or the carbon-induced background (from CH$_2$ target itself) which was not measured and subtracted accordingly;
(iii) the 1$^-$ assignment for the 6.15-MeV state was wrong in Ref.~\cite{bib:gom01}. If their data were correct, the results show
that the 6.15-MeV state most probably has a 3$^-$ or 2$^-$ assignment, while the 6.30-MeV state is the key 1$^-$
state~\cite{bib:arX}. In addition, the inelastic branches of $^{17}$F($p$,$p^\prime$)$^{17}$F$^\ast$ (not measured in
Ref.~\cite{bib:gom01}) can contribute to the $^{14}$O($\alpha$,$p$)$^{17}$F reaction rate considerably. Constraining the
proton-decay branches to the ground and first excited ($E_x$=495 keV, $J^\pi$=1/2$^+$) states of $^{17}$F is therefore of critical
importance. Previously, the inelastic channels were observed for several $^{18}$Ne
levels~\cite{bib:bla03,bib:cern,bib:not04,bib:bar12,bib:alm12}, however, there are still some controversies~\cite{bib:for12}.

We have performed a $^{17}$F+$p$ resonant elastic scattering measurement in inverse kinematics with a $^{17}$F radioactive ion (RI)
beam. The thick-target method~\cite{bib:dae64,bib:art90,bib:gal91,bib:axe96,bib:kub01}, which proved to be a successful
technique in our previous studies~\cite{bib:ter03,bib:ter07,bib:hjj07,bib:yam09,bib:he09,bib:korea,bib:yam13}, was used in this experiment.
This Letter reports our new experimental results. We have resolved the issue with the $J^\pi$ of the 6.15-MeV state and confirmed the 1$^-$
assignment. The resonant properties for other high-lying states were determined and discussed. A new rate of $^{14}$O($\alpha$,$p$)$^{17}$F
has been determined with our results, and its astrophysical impact was examined within the framework of one-zone XRB
postprocessing calculations.

The experiment was performed using the CNS Radioactive Ion Beam separator (CRIB)~\cite{bib:yan05,bib:kub02}, installed by the
Center for Nuclear Study (CNS), the University of Tokyo, in the RI Beam Factory of RIKEN Nishina Center. A primary beam of
$^{16}$O$^{6+}$ was accelerated up to 6.6 MeV/nucleon by an AVF cyclotron ($K$=70) with an average intensity of 560 enA. The
primary beam delivered to CRIB bombarded a liquid-nitrogen-cooled D$_{2}$ gas target ($\sim$90 K)~\cite{bib:yam08} where $^{17}$F
RI beam was produced via the $^{16}$O($d$,$n$)$^{17}$F reaction in inverse kinematics. The D$_{2}$ gas at 120 Torr pressure was
confined in a 80-mm long cell with two 2.5 ${\mu}$m thick Havar foils. The $^{17}$F beam was separated by the CRIB. The $^{17}$F
beam, with a mean energy of 61.9$\pm$0.5 MeV (measured by a silicon detector) and an average intensity of 2.5${\times}$10$^{5}$ pps,
bombarded a thick H$_{2}$ gas target in a scattering chamber located at the final focal plane (F3); the beam was stopped
completely in this target.

\begin{figure}
\begin{center}
\includegraphics[width=7cm]{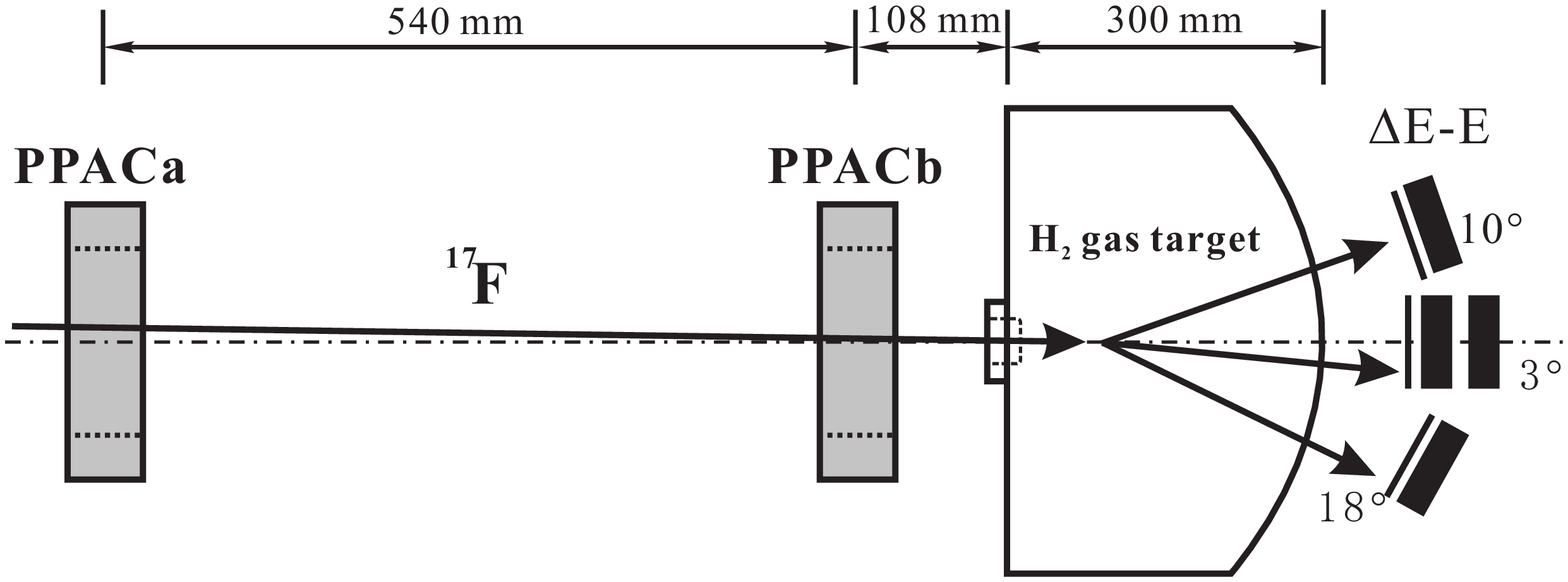}
\end{center}
\vspace{-6mm}
\caption{\label{fig1} Schematic diagram of the experimental setup at the scattering chamber, similar to that used in Ref.~\cite{bib:korea}.
The dotted-dashed line represents the beam axis.}
\end{figure}

\begin{figure}
\includegraphics[width=7cm]{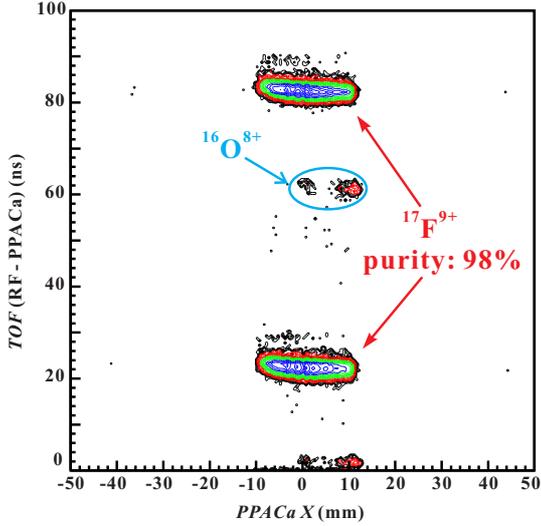}
\vspace{-3mm}
\caption{\label{fig2} (Color online) (a) Identification plot for the beam particles before H$_2$ target via time-of-flight (TOF) technique. Two groups
of particles appear for a single beam, since the data for two extraction cycles of the cyclotron are plotted together. (b) Identification plot
for the recoiled particles via $\Delta E-E$ technique.}
\end{figure}

The experimental setup at the F3 chamber is shown in Fig.~\ref{fig1}, which is quite similar to that used in
Ref.~\cite{bib:korea}.
The beam purity was about 98\% after the Wien-filter. Two PPACs (Parallel Plate Avalanche Counters)~\cite{bib:kum01} provided
the timing and two-dimensional position information of the beam particles. The beam profile on the secondary target was monitored
by the PPACs during the data acquisition. The beam particles were identified event-by-event by the time of flight (TOF) between PPACa
(see Fig.~\ref{fig1}) and the production target using the phase of RF signal provided by the cyclotron. Figure~\ref{fig2}(a) shows the
particle identification at the PPACa. The H$_{2}$ gas target at a pressure of 600 Torr was housed in a 300-mm-radius semi-cylindrical
shape chamber sealed with a 2.5-$\mu$m-thick Havar foil as an entrance window and a 25-$\mu$m-thick aluminized Mylar foil as an
exit window. Comparing to the widely-used solid CH$_{2}$ target, the gas target is free from intrinsic background from carbon.

The recoiling light particles were measured by using three ${\Delta}$E-E Si telescopes at average angles of
$\theta$$_{lab}$$\approx$3$^\circ$, 10$^\circ$ and 18$^\circ$, respectively. In the {\it c.m.} frame of elastic scattering, the
corresponding scattering angles are $\theta_{c.m.}$$\approx$155$^\circ$$\pm$18$^\circ$, 138$^\circ$$\pm$22$^\circ$ and
120$^\circ$$\pm$22$^\circ$, respectively. At $\theta$$_{lab}$$\approx$3$^\circ$, the telescope consisted of a 65-$\mu$m-thick
double-sided-strip (16$\times$16 strips) silicon detector and two 1500-$\mu$m-thick pad detectors. The last pad detector was used
to veto any energetic light ions produced in the production target and satisfying the $B\rho$ selection, possibly not rejected
entirely by the Wien filter because of scattering in the inner wall of the beam line. The configuration of the other two
telescopes is similar to that at $\theta$$_{lab}$$\approx$3$^\circ$, except for the absence of the third veto layer. The position
sensitive $\Delta$E detectors measured the energy, position and timing signals of the particles, and the pad E detectors measured
their residual energies. The recoiling particles were clearly identified by using a $\Delta E-E$ method as shown in
Fig.~\ref{fig2}(b). The energy calibration for the silicon detectors was performed by using a standard triple ${\alpha}$ source
and secondary proton beams at several energy points produced with CRIB during calibration runs. The contribution of background was
evaluated through a separate run with Ar gas at 120 Torr in the target chamber.

\begin{figure}
\includegraphics[width=8.8cm]{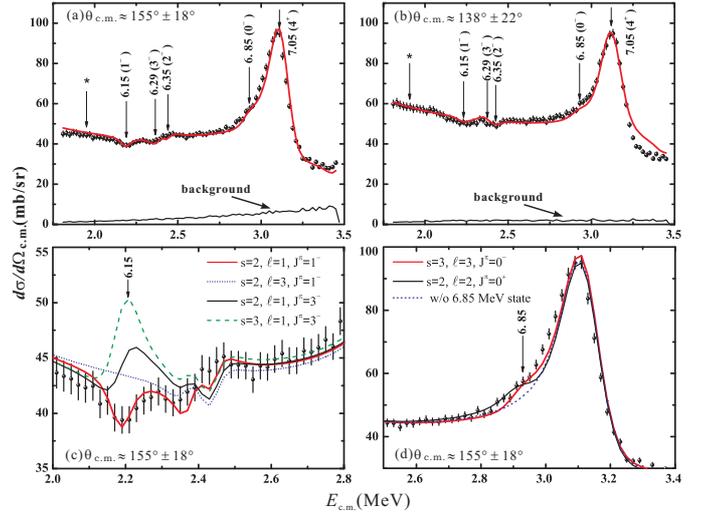}
\vspace{-6mm}
\caption{\label{fig3} (Color online) The center-of-mass differential cross-sections for elastically scattered protons of $^{17}$F+$p$
at angles of (a)$\theta$$_{c.m.}$$\approx$155$^\circ$$\pm$18$^\circ$, and
(b)$\theta$$_{c.m.}$$\approx$138$^\circ$$\pm$22$^\circ$. The (red) curved lines represent the best overall $R$-matrix fits. The
locations of inelastic scattering events for the 6.15-MeV state are indicated as the asterisks. The indicated background spectra (from the Ar
gas run) was subtracted accordingly. Additional $R$-matrix fits for the 6.15- and 6.85-MeV states are shown in (c) and (d),
respectively. See text for details.}
\end{figure}

The excitation functions of $^{17}$F+$p$ elastic scattering have been reconstructed using the procedure described
previously~\cite{bib:he11,bib:kub01,bib:hjj07}. The excitation functions at two scattering angles are shown in Fig.~\ref{fig3}.
The normalized background spectra (taken from the Ar gas run) shown was subtracted accordingly. That of the third telescope (at
$\theta$$_{lab}$$\approx$18$^\circ$) is not shown here due to its worse resolution. Our results demonstrate that the pure H$_{2}$ gas
target allows us to minimize the background protons. It can be regarded as a strong merit comparing to the generally used CH$_2$
solid target which contributes significantly more background from C atoms. The length of the gas target (300 mm) led to an uncertainty
of about 3\% in the solid angle, as determined in event-by-event mode. Such uncertainty in the cross-section is comparable to the
statistical one ($\approx$1\%).

Several resonant structures were clearly observed in the spectra. In order to determine the resonant parameters of observed
resonances, multichannel $R$-matrix calculations~\cite{bib:lan58,bib:des03,bib:bru02} (see examples~\cite{bib:arX,bib:mur09})
that include the energies, widths, spins, angular momenta, and interference sign for each candidate resonance have been performed
in the present work. A channel radius of $R$=1.25$\times$(1+17$^{\frac{1}{3}}$)$\approx$4.46 fm appropriate for the $^{17}$F+$p$
system~\cite{bib:hah96,bib:gom01,bib:he11,bib:wie87,bib:arX,bib:nel85} has been utilized in the calculation. The choice of radius
only has minor effect on the large uncertainties quoted both for the excitation energies and widths.

The ground-state spin-parity configurations of $^{17}$F and proton are $5/2^{+}$ and $1/2^{+}$, respectively. Thus, there are two
channel spins in the elastic channel, {\it i.e.}, $s$=2 and 3. In the present $R$-matrix calculation, the $\alpha$ partial widths
($\Gamma_\alpha$) are neglected relative to the proton widths ($\Gamma_\alpha$$\ll$$\Gamma_p$)~\cite{bib:hah96,bib:har02}. Five
resonances, at $E_{x}$=6.15, 6.28, 6.35, 6.85, and 7.05 MeV, have been analyzed, and the best overall fitting curves are shown in
Fig.~\ref{fig3}(a)\&(b). The resonant parameters obtained are listed in Table~\ref{table1}. In order to fit the data around
$E_{c.m.}$=3.2 MeV, it was necessary to include an additional known resonance ($E_x$$\sim$7.40 MeV, $J^\pi$=2$^+$,
$\Gamma_p$=40 keV)~\cite{bib:he11,bib:har02,bib:for00} in the calculations (see below).

\vspace{2mm}
\noindent\emph{(a) States between 6.1--6.4 MeV}

According to the $R$-matrix analysis, a dip-like structure around $E_{c.m.}$=2.21 MeV, corresponding to the 6.15-MeV state in
$^{18}$Ne, is best fit as a natural-parity 1$^-$ state ($\ell$=1, $s$=2, $\Gamma_p$=50$\pm$15 keV) (see Fig.~\ref{fig3}(c)).
Considering the inelastic branch, this width should correspond to the total width $\Gamma$, and agrees with
$\Gamma$=53.7$\pm$2.6 keV reported before~\cite{bib:bar12}. The resonance shape of this state agrees with that of the
low-statistics experiment by Bardayan {\it et al.}~\cite{bib:bar12}. The natural-parity character of state was also verified by
the previous direct $^{14}$O($\alpha$,$p$)$^{17}$F experiment~\cite{bib:not04}. In addition, as shown in Fig.~\ref{fig3}(c), the
3$^-$ assignment is very unlikely, and also because of the large inelastic branch observed for this state; the unnatural-parity
2$^-$ assignment is also unlikely based on the discussions of the $2p$-emission from this state~\cite{bib:gom01,bib:gri02}.
Therefore, we confirmed the 1$^-$ assignment of the important 6.15-MeV state.
Our resonance shape is entirely different from the bump-like shape observed in Ref.~\cite{bib:gom01}. This may be due to issues in
the data as well as the $R$-matrix analysis (see the lower panel of Fig.~2 in Ref.~\cite{bib:gom01}). As a result, $J^\pi$
assignments suggested in Ref.~\cite{bib:arX} are also questionable.

A structure at $E_{x}$=6.28 MeV was observed in the excitation function, and its shape is reproduced with those resonant
parameters from the work of Hahn {\it et al.}, {\it i.e.}, $E_{c.m.}$=2.36 MeV, $J^\pi$=3$^-$, and $\Gamma_p$=20 keV. In
Ref.~\cite{bib:gom01}, this state was not involved in their $R$-matrix fit. This natural-parity
state was clearly observed in the direct $^{14}$O($\alpha$,$p$)$^{17}$F experiment~\cite{bib:not04}.

The 6.35-MeV state is fitted well with parameters of $J^\pi$=2$^-$, and $\Gamma_p$=10$\pm$5 keV. This $J^\pi$ assignment is
consistent with that speculated by Hahn {\it et al.} It was only weakly populated in the transfer reactions of
($^3$He,$n$) and ($p$,$t$), and unobserved in the direct $^{14}$O($\alpha$,$p$)$^{17}$F experiment~\cite{bib:not04}. With an
unnatural-parity 2$^-$ assignment, this state does not contribute to the rate~\cite{bib:hah96,bib:har02}.

In summary up to this point, we have made confirmation of the three states between 6.1 and 6.4 MeV for the first time, which
has been a long standing problem~\cite{bib:hah96,bib:park}. Because of nuclear structure (4$p$-2$h$ configuration of $h$ (hole)
being in $1p$3/2 and $p$ (particle) in $2s$1/2 or $1d$3/2 orbits), 1$^-$ has very small ($p$,$t$) cross section, and that is why
the 6.15-MeV state was not observed in the previous experiments~\cite{bib:hah96,bib:park}. On the other hand, the 2$^-$ state can
be expected to have appreciable amplitude with a simple $p$-$h$ component, since there is always ($p$,$t$) multistep component
even for an unnatural-parity state~\cite{bib:hah96}. That is why the 6.35-MeV state could be observed even by the ($p$,$t$)
reactions~\cite{bib:hah96,bib:park}; but this 2$^-$ amplitude is significantly smaller than that of 3$^-$ natural-parity state at
6.286-MeV.

The first study to observe inelastic scattering from the 6.15-MeV state was reported by Blackmon {\it et al.}~\cite{bib:bla03}.
They yielded a branching ratio of $\Gamma_{p\prime}$/$\Gamma_{p}$=2.4, and $\Gamma_\mathrm{tot}$$\sim$58 keV, where $\Gamma_{p}$
and $\Gamma_{p\prime}$ are the proton-branching widths for populating the ground and first excited states, respectively. He
{\it et al.}~\cite{bib:cern} detected decay $\gamma$ rays in coincidence with $^{17}$F+$p$ protons looking at the 495-keV $\gamma$
rays, and yielded a ratio of $\Gamma_{p\prime}$/$\Gamma_{p}$$\sim$1. By reanalysis the data in Ref.~\cite{bib:bla03}, Bardayan
{\it et al.}~\cite{bib:bar12} derived a new ratio of $\Gamma_{p\prime}$/$\Gamma_{p}$=0.42$\pm$0.03, and
$\Gamma_\mathrm{tot}$=53.7$\pm$2.0 keV. Most recently, Almaraz-Calderon {\it et al.}~\cite{bib:alm12} populated the 6.15-MeV state
via the $^{16}$O($^3$He,$n$)$^{18}$Ne reaction. Due to large uncertainties, they only estimated the upper limit of this branching
ratio ($\Gamma_{p\prime}$/$\Gamma_{p}$$\leq$0.27). Furthermore, the resolution in the TOF spectrum could result in a relatively
large uncertainty in the excitation energies (see Figure 6 in Ref.~\cite{bib:alm12}). In Fig.~\ref{fig3}(a)\&(b), the position of
the inelastic scattering events is indicated for the 6.15-MeV state. However, no prominent structure was observed for these
inelastic events, and hence the inelastic-scattering channel was not included in the $R$-matrix analysis.

A shell-model calculation for A=17 and 18 nuclides has been performed with a shell-model code OXBASH~\cite{bib:bro92}. The
calculation was carried out in a full model space (spsdpf) using an isospin-conserving WBB interaction of Warburton and
Brown~\cite{bib:war92}. The energy of the second 1$^-$ state was predicted to be $E_x$=6.652 MeV for $^{18}$Ne and $^{18}$O. According to the knowledge of
the mirror $^{18}$O~\cite{bib:li76}, this 1$^-$ state originates mainly from the valence hole of $1p_{3/2}$. The spectroscopic
factors are calculated to be about $S_p(1p_{3/2})$=0.01 for both proton decays to the ground and the first excited states in
$^{17}$F. The calculated value of $S$ is about three times smaller than the experimental one~\cite{bib:li76} in $^{18}$O. Due the
complicated configuration mixing, the theoretical value may fail to reproduce the absolute experimental $S$ value, but the
spectroscopic factor ratio between the ground and first excited state should be reliable. The calculated branching ratio is $\Gamma_{p\prime}$/$\Gamma_{p}$$\approx$0.66 with a partial proton width relation of
$\Gamma_{p}$=$\frac{3\hbar^2}{\mu R^2}$$P_{\ell}$$C^2S_{p}$~\cite{bib:wie87}. The calculated proton width is about
20 keV with $C^2S_{p}$=0.01. These results are reasonable given the measurement by Bardayan {\it et al.}~\cite{bib:bar12}

\vspace{2mm}
\noindent\emph{(b) State at 6.85 MeV}

It is very interesting that a shoulder-like structure around $E_{c.m.}$=2.93 MeV was observed by both telescopes as shown in
Fig.~\ref{fig3}(a)\&(b). This is possibly a new state at $E_{x}$=6.85$\pm$0.10 MeV. Both $J^\pi$=0$^-$ or 0$^+$ resonances can
reproduce the observed shape as shown in Fig.~\ref{fig3}(d). Because of the small energy shift for the negative-parity states in
this excitation energy region~\cite{bib:for00}, such a state is possibly the analog state of $^{18}$O at $E_x$=6.880 MeV
(0$^-$)~\cite{bib:til95}. In fact, Wiescher {\it et al.}~\cite{bib:wie87} predicted a $J^\pi$=0$^-$ state in $^{18}$Ne, analog to
the 6.88 MeV state in $^{18}$O, at 6.85 MeV with a proton spectroscopic factor of $C^2S_p$=0.01. However, another possibility
still exists as discussed below.

A strong proton resonance from a state at $E_x$$\sim$6.6 MeV was observed in an earlier direct $^{14}$O($\alpha$,$p$)$^{17}$F
experiment~\cite{bib:not04}. Because no such state was previously observed in $^{18}$Ne, Notani {\it et al.} speculated that it
might be due to a state at $E_x$$\sim$7.1 MeV decaying to the first excited state of $^{17}$F. Later on, a careful $^{17}$F+$p$
scattering experiment~\cite{bib:bar10} was performed, but no evidence of inelastic $^{17}$F+$p$ scattering was observed in this
energy region, and the decay branching ratio to the first excited state ($\Gamma_{p\prime}$/$\Gamma_{p}$) was constrained to be
$<$0.03. Almaraz-Calderon {\it et al.} recently reported a ratio of 0.19$\pm$0.08 for the 7.05 MeV state. Later on, this large
ratio was questioned by Fortune~\cite{bib:for12} who estimated a ratio less than about 2$\times$10$^{-4}$, in agreement with an
earlier limit of $\leq$1/90 from Harss {\it et al.}~\cite{bib:har02}. Based on the suggestion of Fortune, Almaraz-Calderon {\it et al.}
thought that their large number might be attributed from an unknown state at $E_x$$\sim$6.7 MeV in $^{18}$Ne. In fact, there is a
hint of a weak state observed at $E_x$$\sim$6.8 MeV (see Figure 6 in Ref.~\cite{bib:alm12}). As discussed above, such a state at
$E_x$=6.85$\pm$0.10 MeV was also observed in the present work. Therefore, we conclude that very likely a new state around
6.8 MeV exists in $^{18}$Ne. Since this state was populated in the direct $^{14}$O($\alpha$,$p$)$^{17}$F reaction, it should have
a natural parity. Thus, it is also possibly a candidate of the $J^\pi$=0$^+$ state, a bandhead state of the six-particle four-hole (6$p$-4$h$)
band~\cite{bib:for03,bib:for11}. If this 6.85-MeV state were 0$^+$, its $\alpha$ width would be roughly 149 eV, as estimated with the
expression of $\Gamma_{\alpha}$=$\frac{3\hbar^2}{\mu R^2}$$P_{\ell}(E)$$C^2S_\alpha$~\cite{bib:wie87}. Here, a
spectroscopic factor of $C^2S_\alpha$=0.01 were assumed in the calculation. As such, if the state is 0$^+$ ($\omega\gamma$=149 eV),
its contribution to the $^{14}$O($\alpha$,$p$)$^{17}$F rate would be larger than that of the 7.05-MeV state ($\omega\gamma$=203 eV); but
it is still much smaller than that of the 6.15 MeV state below $\sim$2.5 GK.
Of course, if it is, in fact, 0$^-$, it would not contribute at all. The exact $J^\pi$ for this 6.85 MeV state still needs to be
determined by additional experiments (although we prefer a 0$^+$), and hence this state was not involved in our rate calculation.

\vspace{2mm}
\noindent\emph{(b) States at 7.05 and 7.35 MeV}

The state~\cite{bib:har02} at $E_{x}$=7.05 MeV (4$^+$, $\Gamma_p$=95 keV) was also observed at $E_{c.m.}$=3.13 MeV.
However, the doublet structure around $E_{x}$=7.05 and 7.12 MeV suggested in Refs.~\cite{bib:hah96,bib:he11} could not be resolved
within the present energy resolution ($\sim$80 keV in FWHM in this region). A single peak is adequate for the fit to our data, with
similar $\chi$$^2$ value to a fit using two peaks.

One state around 7.35 MeV was observed in the ($^3$He,$n$) and ($^{12}$C,$^6$He) reactions~\cite{bib:hah96} and
showed (1$^-$, 2$^+$) characteristics in the ($^3$He,$n$) angular distribution. Hahn {\em et al.}~\cite{bib:hah96}
suggested a 1$^-$ for this state based on a very simple mirror argument. Later on, following the arguments of Fortune and
Sherr~\cite{bib:for00}, Harss {\em et al.}~\cite{bib:har02} speculated it as a 2$^+$ state based on a Coulomb-shift discussion.
Our present and previous results~\cite{bib:he11} all support the 2$^+$($\ell$=2) assignment. However, its mirror partner is
still uncertain~\cite{bib:for11}. Combining with the discussion of Fortune and Sherr~\cite{bib:for11}, we speculate that a new
7.796-MeV state recently observed~\cite{bib:oer10} in $^{18}$O may be the mirror of the 7.35 MeV state in $^{18}$Ne. This would
imply that the bandhead (0$^+$) of the six-particle four-hole (6$p$-4$h$)~\cite{bib:for03,bib:for11} band in $^{18}$O is still missing.

By evaluating all the available data, the resonance parameters adopted for the $^{14}$O($\alpha$,$p$)$^{17}$F resonant rate
calculations are summarized in Table~\ref{table2}. Similar to the method utilized by Hahn {\it et al.}~\cite{bib:hah96} and
Bardayan {\it et al.}~\cite{bib:bar97}, the $^{14}$O($\alpha$,$p$)$^{17}$F total rate has been numerically calculated using the
resonance parameters listed in Table~\ref{table2} and the direct reaction $S$-factors calculated by Funck \&
Langanke~\cite{bib:fun88}. Here, the interference between the direct-reaction $\ell$=1 partial wave and the 6.15-MeV (1$^{-}$)
excited state was included in the calculations; the inelastic branches (listed in Table~\ref{table2}) were also included in the
integration. Two different $^{14}$O($\alpha$,$p$)$^{17}$F rates~\cite{bib:rate} were calculated by assuming the constructive
(``Present+") and destructive (``Present-") interferences between the direct and resonant captures (for the 6.15-MeV state).
These two rates differ by a factor of $\approx$5 at 0.35 GK and less than 10\% at 1 GK.
In the temperature region of 0.3--2 GK, our ``Present+" rate is about 1.1--2.2 times larger than the corresponding rate
from Hahn {\it et al.} (``Hahn+"), and the ``Present-" rate is a factor of 1.4--2.7 larger than that of ``Hahn-". Our adopted
parameters are more reliable than the older ones determined by Hahn {\it et al.} about twenty years ago. In addition, below 0.3 GK,
our rates are orders of magnitude greater than the rates of Harss {\it et al.}~\cite{bib:har02} and
Alamaraz-Calderon {\it et al.}~\cite{bib:alm12}, which were calculated by using the simple narrow-resonance formulism (without
considering interference effects). Between 0.4 and 2 GK, the ``Present+" rate is a factor of 1.1--2.7 greater than that from
Harss {\it et al.}, and a factor of 1.3--3.2 greater than that of Almaraz-Calderon {\it et al.} In addition, our rates are larger
than the older rate estimated by Wiescher {\it et al.}~\cite{bib:wie87} by factors of $\approx$2--100 over the temperature region
of 0.3--2 GK.

\begin{table}
\caption{\label{table1} Resonant parameters derived from the present $R$-matrix analysis. The excitation energies are the average
values derived from our data sets, and uncertainties are estimated by a Monte-Carlo simulation. The widths available in the
literature are listed for comparison.}
\begin{tabular}{|l|c|c|l|l|}
\hline \hline
$E_x$ (MeV) & $J^\pi$ & $\ell$ & $\Gamma_p$ (keV) & $\Gamma$ (keV)$^\mathrm{literature}$                 \\
\hline
6.15(0.03)  & 1$^{-}$ & 1      &  50(15)$^a$  & $\leq$40~\cite{bib:hah96}; 53.7$\pm$2.0~\cite{bib:bar12} \\
6.28(0.03)  & 3$^{-}$ & 1      &  20(15)      & $\leq$20~\cite{bib:hah96}; 8$\pm$7~\cite{bib:par99}      \\
6.35(0.03)  & 2$^{-}$ & 1      &  10(5)       & 45$\pm$10~\cite{bib:hah96}; 18$\pm$9~\cite{bib:par99}    \\
6.85(0.11)$^b$ & 0$^{-}$ & 3  &  50(30)      &                                                          \\
            & 0$^{+}$ & 2      &  50(30)      &                                                          \\
7.05(0.03)  & 4$^{+}$ & 2      &  95(20)      & $\leq$120~\cite{bib:hah96}; 90$\pm$40~\cite{bib:har02}   \\
\hline \hline
\end{tabular}
\footnotesize
$^a$ This width corresponds to the total width $\Gamma_\mathrm{tot}$ due to the inelastic branch.\\
$^b$ Large uncertainty mainly originates from the $R$-matrix fit. Our data are consistent with either a 0$^+$ or 0$^-$ assignment to this state.
\end{table}

\begin{table*}
\caption{\label{table2}Resonance parameters adopted in the calculation of the $^{14}$O($\alpha$,$p$)$^{17}$F reaction rate.}
\begin{tabular}{|c|c|c|c|c|c|c|c|}
\hline \hline
$E_x$ (MeV) & $E_{res}$ (MeV) & $J^{\pi}$ & $\Gamma_{\alpha}$ (eV) & $\Gamma_{p}$ (keV)   & $\Gamma_{p{\prime}}$ (keV) & $\Gamma$ (keV) & $\omega$$\gamma$ (MeV)\\
\hline
5.153$^a$ & 0.039  & 3$^{-}$  & 4.3$\times$10$^{-52}$$^a$         & 1.7$^a$          &                  & $\leq$15$^a$     & 3.0$\times$10$^{-57}$ \\
6.150$^a$ & 1.036  & 1$^{-}$  & 3.9$\pm$1.0$^b$                   & 37.8$\pm$1.9$^c$ & 15.9$\pm$0.7$^c$ & 53.7$\pm$2.0$^c$ & 1.2$\times$10$^{-5}$  \\
6.286$^a$ & 1.172  & 3$^{-}$  & 0.34$^a$                          & 8$\pm$7          &                  & 8 $\pm$7$^d$     & 2.4$\times$10$^{-6}$  \\
7.05$^a$  & 1.936  & 4$^{+}$  & 22.6$\pm$3.2$^e$                  & 90$\pm$40        &                  & 90 $\pm$40$^f$   & 2.0$\times$10$^{-4}$  \\
7.37$^f$  & 2.256  & 2$^{+}$  & 40$\pm$30$^f$                     & 70$\pm$60        &                  & 70 $\pm$60$^f$   & 2.0$\times$10$^{-4}$  \\
7.60$^f$  & 2.486  & 1$^{-}$  & 1000$\pm$120$^f$                  & 72$\pm$20$^f$    & $<$2$^f$         & 75$\pm$20$^f$    & 3.0$\times$10$^{-3}$  \\
7.95$^g$  & 2.836  & 3$^{-}$  & (11$\pm$6.6)$\times$10$^{3}$$^g$  & 35$\pm$15$^g$    & 9.0$\pm$5.6$^g$  & 55$\pm$20$^g$    & 6.2$\times$10$^{-2}$  \\
8.09$^g$  & 2.976  & 3$^{-}$  & (6.3$\pm$3.9)$\times$10$^{3}$$^g$ & 20$\pm$4$^g$     & 4$\pm$3$^g$      & 30$^a$           & 3.5$\times$10$^{-2}$  \\
\hline \hline
\end{tabular}
\footnotesize \\
$^a$ From Hahn {\it et al.}~\cite{bib:hah96}; $^b$ From Fortune~\cite{bib:for12a}; $^c$ From Bardayan {\it et al.}~\cite{bib:bar12}; \\
$^d$ From Park {\it et al.}~\cite{bib:par99}; $^e$ From Fortune~\cite{bib:for12}; $^f$ From Harss {\it et al.} \cite{bib:har02}; $^g$ From Almaraz-Calderon {\it et al.}~\cite{bib:alm12}.
\end{table*}

\begin{figure}
\includegraphics[width=8cm]{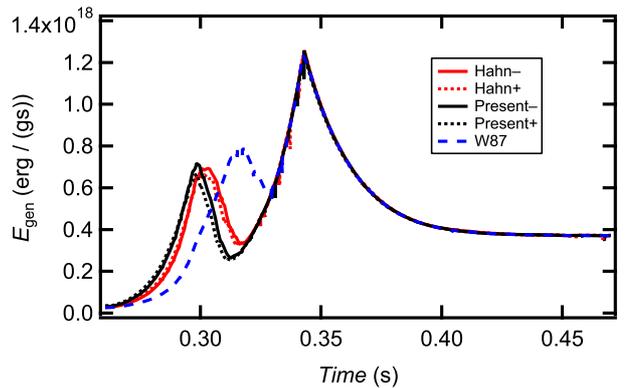}
\vspace{-3mm}
\caption{\label{fig4} (Color online) Nuclear energy generation rates during one-zone XRB calculations using the K04 thermodynamic
history~\cite{bib:par08}. Results using the ``Present" rates (black solid line for destructive ``-", black dotted line for
constructive ``+") and the ``Hahn" rates~\cite{bib:hah96} (red solid line for ``-", red dotted line for ``+") are indicated. The
result using the estimated rate of Wiescher {\it et al.}~\cite{bib:wie87} is also shown for comparison (labeled as ``W87").
See text for details.}
\end{figure}

The impact of these new $^{14}$O($\alpha$,$p$)$^{17}$F rates was examined using one-zone XRB models. With the representative K04
temperature-density-time thermodynamic history ($T_\mathrm{peak}$=1.4 GK~\cite{bib:par08}), the nuclear energy generation rate
($E_\mathrm{gen}$) during the XRBs has been studied by performing separate post-processing calculations with seven different rates:
two present rates (``Present+"~\&~``Present-"), as well as previous rates from Wiescher {\it et al.}~\cite{bib:wie87},
Hahn {\it et al.}~\cite{bib:hah96} (``Hahn+"~\&~``Hahn-"), Harss {\it et al.}~\cite{bib:har02}, and Alamaraz-Calderon {\it et al.}~\cite{bib:alm12}.
Figure~\ref{fig4} shows the differences in $E_\mathrm{gen}$ at early times of the burst, as calculated using the
``Present+"~\&~``Present-", ``Hahn+"~\&~``Hahn-" and Wiescher {\it et al.}~\cite{bib:wie87} rates. It shows that the shape and
time structure of $E_\mathrm{gen}$ are influenced considerably by the rates. For example, at about $\approx$0.31 s relative to the
start of the burst, the ``Present+" rate gives an $E_\mathrm{gen}$ that is a factor of $\approx$1.5 less than that from the ``Hahn+"
rate and a factor of $\approx$3 less than that from ``W87". Note that the sign of the interference only has a marginal ($<$10\%)
effect on the predicted $E_\mathrm{gen}$. The predicted $E_\mathrm{gen}$ profiles using the reaction rates of
Harss {\it et al.} and Almaraz-Calderon {\it et al.} are not shown
in Fig.~\ref{fig4}. These profiles differ from that of the ``Present-" profile only between $\approx$0.30--0.32 s,
where they lie between the ``Present-" and ``Hahn-" curves. Given the key role of the $^{14}$O($\alpha$,$p$)$^{17}$F reaction in
the breakout from the HCNO cycle during an XRB, it is precisely at early times (low T) where different rate could be expected
to affect the nuclear energy generation. These results are also in accord with results from
Ref.~\cite{bib:par08} where variations of the $^{14}$O($\alpha$,$p$)$^{17}$F rate by a (uniform) factor of ten were found to
significantly affect $E_\mathrm{gen}$ in the K04 model.

The nuclear energy generation rate predicted in the adopted one-zone XRB model is well-constrained by our new reaction rates and
differs from $E_\mathrm{gen}$ predictions using previous $^{14}$O($\alpha$,$p$)$^{17}$F rates. As such, reaction rate
libraries~\cite{bib:jina,bib:ornl} incorporating older $^{14}$O($\alpha$,$p$)$^{17}$F rates should be updated. Additional tests
using detailed hydrodynamic XRB models should be performed to confirm these results and examine the impact of different
$^{14}$O($\alpha$,$p$)$^{17}$F rates further.

\vspace{0.5cm}
\textbf{Acknowledgments}
\vspace{0.2cm}

We would like to thank the RIKEN and CNS staff for their kind operation of the AVF cyclotron. This work is financially supported by the NNSF of China
(Nos. 11135005, 11321064), the 973 Program of China (2013CB834406), as well as supported by JSPS KAKENHI (No. 25800125). A.P was supported by the Spanish
MICINN (Nos. AYA2010-15685, EUI2009-04167), by the E.U. FEDER funds as well
as by the ESF EUROCORES Program EuroGENESIS. A.A.C and D.I were supported by the National Science and Engineering Research Council of Canada. K.I.H was
supported by the NRF grant funded by the Korea government (MSIP) (No. NRF-2012M7A1A2055625), and J.Y.M, H.S.J, and C.S.L by the Priority Centers Research
Program in Korea (2009-0093817).


\begin{thebibliography}{99}
\bibitem{bib:rate}
The present (+/-) recommended rates can be analytically expressed in terms of $T_9$ by $N_{A}\langle\sigma v\rangle^+_{-}$=$\mathrm{exp} [^{1.890}_{1.301}\times 10^2 - ^{0.179}_{0.157} T_9^{-1} - ^{4.872}_{10.731} T_9^{-1/3}
 - ^{2.358}_{1.445}\times 10^2 T_9^{1/3} + ^{74.11}_{18.51} T_9 - ^{22.61}_{1.67} T_9^{5/3}
+ ^{52.71}_{38.83} \mathrm{ln}T_9] + \mathrm{exp} [- ^{2.149}_{1.677}\times 10^2 + ^{11.843}_{10.333} T_9^{-1}
 - ^{9.092}_{7.892}\times 10^2 T_9^{-1/3}
 - ^{1.166}_{0.994}\times 10^3 T_9^{1/3} - ^{56.26}_{49.47} T_9 + ^{26.82}_{24.73} T_9^{5/3} - ^{6.179}_{5.277}\times 10^2 \mathrm{ln}T_9]$.
\end{thebibliography}
\end{document}